\documentclass[twocolumn]{aastex6}

\usepackage{amsmath}
\usepackage{soul}  

\hypersetup{
    pdffitwindow=false,            
    pdfstartview={Fit},            
    pdftitle={Spiral_disk_instability},     
    pdfauthor={Rahul Kashyap},         
    pdfsubject={Astrophysics},                 
    pdfnewwindow=true,             
}

\begin{document}
	
\title{The Differences between Analytical and Numerical Ignition Curve of He-C-O Mixture}

\author{Rahul Kashyap\altaffilmark{1}}
\altaffiltext{1} {Physics Department, Pennsylvania State University, State College, PA, USA}

\begin{abstract}
	
In this paper, we use large reaction networks to find ignition conditions of single-zone nuclear fuel with compositions typical of white dwarf (WD) matter. The necessary but, possibly not sufficient condition for initiation of detonation is that the nuclear burning proceeds on smaller timescale than the sound-crossing timescale. Under typical white dwarf thermodynamic conditions, the nuclear timescale depends sensitively on the chosen nuclear reaction network, and widely-used analytical formulae are not sufficient to accurately determine unstable ignition conditions the answer to which lies in the fully compressible reactive turbulent simulations. We model the nuclear reactions in the numerical simulation using the astrophysical code TORCH. We report the differences introduced due to the size of the nuclear reaction network. Our findings have implications for numerical results obtained using the popular 19-species network in multidimensional simulations of WD of Chandrasekhar and sub-Chandrasekhar masses. We present a sufficiently accurate general criterion in temperature-density ($T-\rho$) plane for a given initial fuel to ignite in an unstable fashion.

\end{abstract}


\section{Introduction} \label{sec:intro}

Nuclear reactions play an important role in the evolution of stars as well as formation of astrophysical transients such as X-ray bursts, novae and supernovae. The transient phenomenon in astrophysics indicates the burning of nuclear fuel at much shorter timescale than the typical life time of their progenitors. In fact, such timescales also has to be smaller than the dynamical timescales of these systems. In particular, Type Ia supernovae (SNeIa) produces $10^{51}$ ergs which equivalent to the energy produced by a low-mass star like our Sun in $\sim 10$ billion years. What are the precise general condition that makes a nuclear fuel burn so fast in order to produce large amount of energy in a short time? It is one of the most important unsolved question in theoretical and computational modeling of astrophysical flows. 

SNe Ia forms from the explosion of white dwarfs when C/O material reaches explosive condition \citep{Hoyle-Fowler1960A} . SNe Ia are one of the most important tools in modern cosmology \citep{Riess1998A}. The observables of these events have characteristic behaviors unique to SNe Ia. This allows us to calibrate SNe Ia to measure the distances and redshifts of their host galaxies. The empirical relation \citep{Phillips1993A} used for this purpose has been verified for nearby SNe Ia using independent means. However, the fundamental physics connecting the progenitor conditions to their observable properties are not yet fully understood. In this work we focus on the problem of detonation in white dwarfs in order to shed light on SNe Ia progenitor problem. 

Two main channels have been proposed in the literature to have suitable conditions for carbon detonation in WD -- Single degenerate and Double Degenerate systems, depending on whether system has one or two white dwarfs. In addition to them, a thin He layer on a white dwarf is a natural outcome of the stellar evolution. It is easier for He to reach detonation condition than C during the final phase of binary WD mergers. The He-detonation, in turn, can cause carbon-detonation through the focussing and convergence of shock detonation wave in the degenerate material \citep{nomoto82,Woosley1994A,Livne1990A,Livne1995A}. This hypothesis, although interesting, has not been shown to work in a numerical simulation thus far except in one and two dimensional models where the focussing of shock is geometric and occurs due to the assumption of symmetry. Here we discuss one of the important consequences of mixing between between He layer and C/O core.\\

In the numerical simulations presented in the literature corresponding to the above proposed mechanisms, the full 3D simultaions of a turbulent reactive flow have not been achieved primarily due to immensity of computational cost. This makes the question of transition to detonation very acute which has been mostly introduced artificially \citep{Gamezo2004}. Recently, \citealt{Zenati2020} studied the effects of turbulent dissipation on the initiation of detonation. They determine the condition of initiation of detonation for the reaction, $\mathrm{C}^{12}(\mathrm{O}^{16},\alpha)\mathrm{Mg}^{24}$ and the critical length scale at which the nuclear timescale becomes smaller than sound-crossing timescale.

We address here an indpendent part of the problem of detonation which is -- How does the ratio of nuclear energy generation timescale and the dynamical timescale depend upon the size of nuclear reaction network. The initiation of detonation is facilitated when this ratio becomes smaller than one. We emphasie that our study must be complemented with the determination of the critical lengthscale required for detonation where the effects presented in this work are most pronounced. We present how a mixed fuel carbon/oxygen in Helium or, He in C/O material enhances the nuclear energy produced. Both types of conditions are very natural in the context of sub-Chandrasekhar mass binary white dwarfs. 

\subsection{Initiation and Propagation of Self-sustained Detonation}
The combustion of degenerate matter is required to proceed  at supersonic speed in the form of detonation wave in order to produce enough energy which would unbind the star to form SNe Ia. This is also necessary to bring the reaction to equillibrium which would produce enough $^{56}$Ni to power the observed peak brightnesses and light curves. However, this is a strict requirement only for the significantly sub-Chandrasekhar mass white dwarfs. Some of the deflagration models in single degenerate model might be responsible for subluminous SNe Ia although not all classes.\\

The self-sustained detonation wave requires a shock wave followed by a reaction zone in Zel'dovich-von Neumann-Doering (ZND) picture of detonation wave. The thickness of the reaction zone is determined by the nuclear reaction timescale and hence also by the strength of the shock.\\

One of the important problem in SNe Ia modelling is to obtain the physically realistic picture of initiation of detonation in degenerate matter. Detonation has been observed to be initiated in terrestrial chemical combustion via two means -- spontaneous or, prompt detonation and deflagration-to-detonation transition. One would naturally assume, this to be the same in case of supernovae owing to the fact that the physical model of both, terrestrial and astrophysical combustion bears much similarity.\\

These two mechanisms has been explored extensively in the case of Chandrasekhar mass white dwarfs \cite{khokhlov91}. The prompt detonation scenario overproduce iron-group elements and hence cannot be the explanation for all SNe Ia. The deflagration-to-detonation transition \citep{Khokhlov1997} requires careful tuning of the transition density at which subsonic deflagration wave becomes supersonic detonation wave. The existence and parameter governing both scenarios of detonation are largely unphysical to this date. In this paper, we will present the investigations into conditions for unstable ignition in sub-Chandrasekhar mass WDs. However, the methodology unnnderlying the arguments apply to a wide class of nuclear matter such as novae, core-collapse supernovae and other such explosive environment.

\cite{Seitenzahl_2009} has attempted to answer this question in one dimensional simulations using 13-isotope network -- Which of the given initial temperature profile, peak temperature and an ambient temperature and ambient density would develop a supersonic detonation wave. This development could be prompt or gradual which would depend on the details. One can vary these four conditions to produce the detonation conditions in a wide variety of situations. However, the feasibility of the initial conditions must be examined. The milder the conditions which produce detonation wave, the more probable it would be. Additionally, the whole analysis should also be done in three dimensions for determining accurate criterion. In the present study, we expect to hint at the effects of large reaction network in such studies. 

Here, first in Section \ref{nuclear_numerical}, we present the numerical methods used in astrophysical simulations, the method employed to solve the reaction network, and the coupling scheme.  We then describe our results and implications of our results on computational results in literature.

\section{Numerical Integration of 46-species Nuclear Reaction Network} \label{nuclear_numerical}

We use the publicly available code TORCH for solving nuclear reaction networks. We establish a simple local ignition criterion using three conditions together:  $\dot{\epsilon } > 0$,  $\frac{e_{\rm int}}{\dot{\epsilon}}<t_{\rm dyn}$, $t_{\rm burn,i} = t_{\rm dyn}$. Here $\dot{\epsilon }$ is nuclear energy generation rate (after subtracting neutrino losses), $e_{\rm int}$ is specific internal energy of the cell. Here we take the dynamical time to be the free-fall time $t_{\rm dyn} = (G \rho)^{-1/2}$ where $\rho$ is density of the cell. We emphasize that any other mixing timescale including the sound-crossing time would be longer than the free-fall time. Hence, we provide a conservative limit for burning condition in this sense. Burning timescale of element $i$ is defined as $t_{\rm burn,i}=X_{i}/ \dot{X_{i}}$ where $X_{i}$ is the mass fraction.

The first two conditions ensure that the reaction is injecting energy without allowing the medium to respond hydrodynamically, while the third condition tells us which of the species is burning primarily to drive the condition towards ignition. We develop two estimates, for helium and carbon ignition respectively, following these criteria.

Following other authors \citep{Dan2014A}, we develop an analytic form for the burning timescale $t_{\rm burn}$ for both carbon and helium burning using  single-reaction rates. For the case of carbon burning, we use the reaction rate for  $^{12}$C ($^{12}$C,$\alpha$) $^{20}$Ne \citep{Blinnikov1987A,Fowler1988A}. For the case of helium burning, we use the reaction rate for triple $\alpha$ \citep{Kippenhahn1990A,Fowler1988A}. Secondly, we also calculate the burning timescale $t_{\rm burn}$ using an extended reaction network with 19, 46 and 204 isotopes repectively using the publicly available code TORCH \citep{Timmes1999}. We numerically integrate the initial composition for an array of temperature and density under isobaric conditions. The isobaric condition is representative of the assumption of local thermodynamic equillibrium (LTE) approximation in the interior of the whitedwarfs.  Different initial compositions are used for carbon (50-50 C/O) and helium burning (pure He). For each initial temperature, the abundances are advanced up to one dynamical time. We find the density at which all three conditions are satisfied together. The locus of these points yields the desired detonation condition as a curve in the density-temperature plane.

The molar mass fraction of species at a time $t$, $Y_i(t)$ is then integrated over a time step (one dynamical time) to obtain the final abundances of species $Y_{i}(t+\Delta t)$ after time $\Delta t$. The specific nuclear energy yield (we will call it $\dot{s}_{bin}$ in erg gm$^{-1}$) is calculated using Einstein's mass-energy relation:
\begin{align}
\dot{s}_{bin} =N_A \sum\limits_{i} B_i\frac{[{Y_i(t+\Delta t)-Y_i(t)}]}{\Delta t}
\end{align}
where $B_i$ is the binding energy of $i$th species and $N_A$ is the Avogadro number. The molar mass fraction, $Y_i(t) (= \frac{n_i}{\rho N_A})$ is defined in terms of mass fraction, $X_i$, total baryon mass density, $\rho$ and number density of $i^{th}$ isotope, $n_i$. Some fraction of this energy is produced in the form of kinetic energy of neutrinos. Hence, the net output of energy responsible for hydrodynamic evolution is obtained after subtracting the energy carried away by the neutrinos produced during nuclear reactions\citep{Itoh_1996}. We get the total energy output after subtracting specific neutrino losses $\dot {s}_{\rm neut}$ from the specific mass defect energy $\dot {s}_{\rm bin}$.
\begin{align}
\frac{d \epsilon}{dt} & = \dot{\epsilon} = \dot{s}_{bin} - \dot{s}_{neut}
\end{align}

Smaller reaction networks are approximations to larger, more realistic networks with a choice to make the energetics more accurate for hydrodynamical calculations. The Approx13 and Approx19 networks in FLASH can reproduce the energy generation rate of larger networks to within 30\% \citep{Timmes_2000}. However, the accuracy is significantly reduced in the case of heavy neutronization. Few important questions arises in the context of SNe Ia. Does the use of a limited nuclear reaction network such as Approx19 curtail critical reaction pathways permissible in larger reaction networks? Can the use of a larger network enable unstable nuclear burning which are suppressed by smaller networks?\\

\section{Results}

We determine the ignition conditions for any given initial fuel composition. In particular, we present the ignition condition for C/O mixed with He and He mixed with C/O. These compositions are more realistic and play an important role in the problem of type Ia supernovae progenitors. We emphasize that a successful detonation also requires a minimum length-scale of the fuel to have the above mentioned conditions \citep{Seitenzahl_etal_2009}. We have not studied the length requirements in light of our findings here. However, we expect the previously found conditions to become somewhat weaker because of enhanced nuclear energy production due to the use of larger reaction networks and turbulent dissipation.

\subsection{Local Criteria for Self-Sustained Ignition}

Our criteria for He and C ignitions including full reaction networks differ markedly from the previous results, such as that of \cite{Dan2014A}, where analytical results based only on 3-$\alpha$ reactions have been used. The extended nuclear reaction network gives particularly different results in the regime $T>8.7 \times 10^8$~K and $\rho< 8.3 \times 10^4$~g~cm$^{-3}$ where the approximations of the 3-$\alpha$ analytical calculations break down. Our results are similar to the findings of \cite{Shen2014A}, who have used similar reaction networks, and with a more reasonable temperature profile.\\

\subsubsection{Ignition in Helium: pure and mixed with Carbon}
The question of carbon detonation in white dwarfs can also be investigated via another, closely-related question, which is: under what conditions does the presence of a small surface layer of helium ignite? An attractive alternative detonation mechanism to direct carbon ignition involves igniting a thin surface layer of helium atop the WD. The detonation of a thick helium layer atop a WD core transmits a diverging detonation outward and a converging detonation inward \citep{nomoto82,Woosley1994A,Livne1990A,Livne1995A}. The converging inwardly-directed detonation in turn causes a shock wave propagation into the C/O WD core, possibly leading to a detonation either on the helium-carbon interface (an ``edge-lit detonation''), or at the center due to shock convergence. However, these studies depend crucially upon the geometry and the shock convergence could be purely geometrical effect resulting from the assumptions of symmetry. A variation on this mechanism invokes a dynamically-ignited helium layer \citep{Guillochon2010A, Pakmor_2012}. 

Recent results \citep{Shen2014A} have shown that the mass of the He-shell required may be much smaller ($\sim 10^{-2}\, M_{\sun}$) than previously thought, if one includes the effect of $\alpha$-captures in the He-shell contaminated with an admixture of carbon and oxygen to act as seed nuclei. The significance of these $\alpha$-captures is not captured in the smaller approximate network reactions frequently employed in multidimensional simulations, but can be accommodated in larger reaction networks.\\

\begin{figure}
	\begin{center}
		\includegraphics[trim=10 0 10 10,clip,width=0.9\columnwidth]{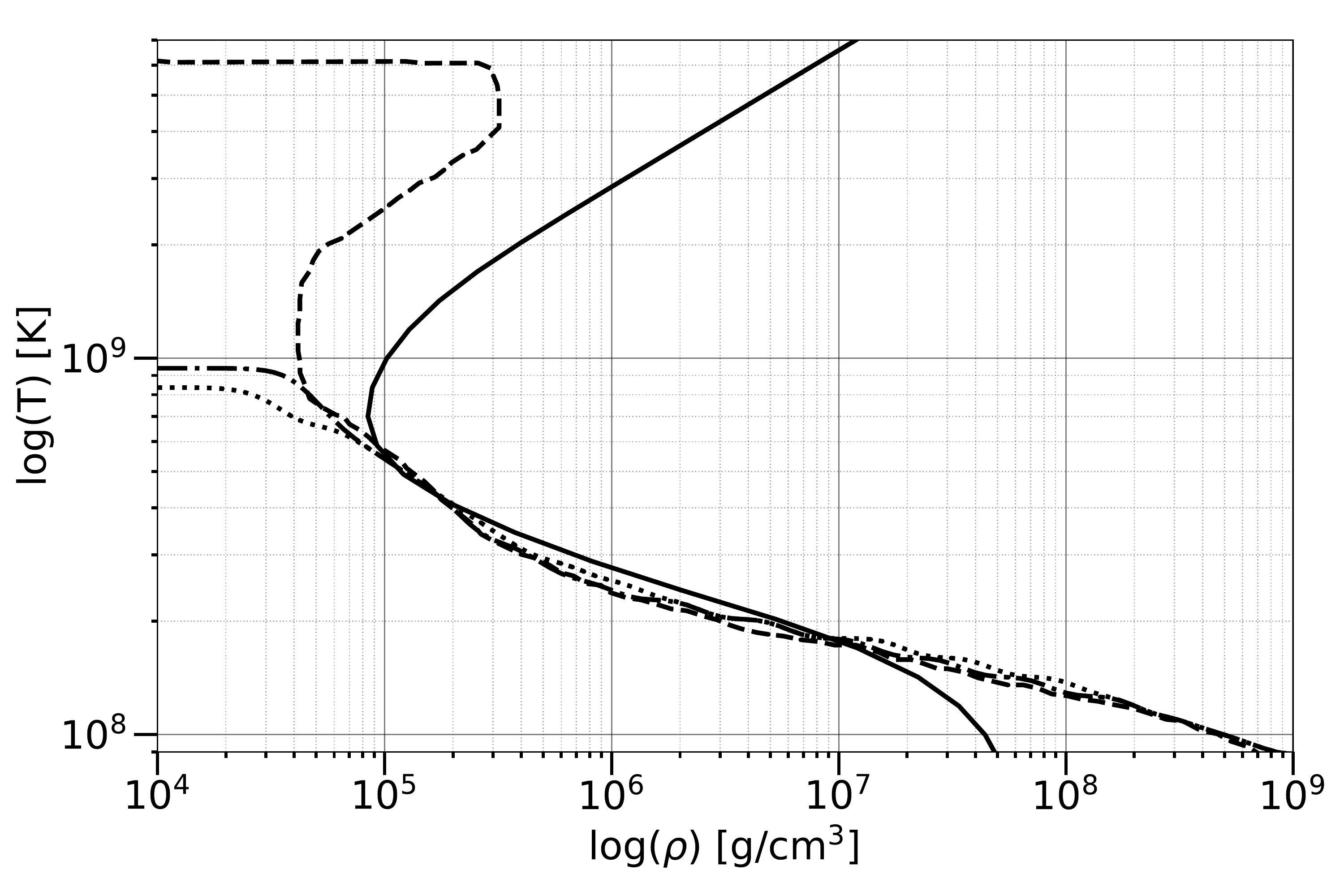}
		\caption{Local ignition criteria in $T-\rho -$plane for fuels containing Helium mixed with small amount of Carbon. The solid black curve represent the analytical $t_{dyn}/t_{nuc}=1$ curve for triple-$\alpha$ Helium ignition curve using analytical results. The dashed black curve represent the ignition curve for pure Helium using 46 isotopes. The dot-dashed and dotted curves represent the ignition curves where the initial fuel is taken to be 95/2.5/2.5 and 80/10/10 for He/C/O respectively.}
		\label{helium46}
	\end{center}
\end{figure}

In figure \ref{helium46}, we compare the analytical ignition curve for pure Helium with the one obtained from 19 and 46 species reaction network. For the analytical curve, we assume that all $^{4}$He fuses via only one pathway -- 3$\alpha$ to form $^{12}$C only. In a realistic situation, after $^{12}$C is produced from $^{4}$He, $^{12}$C will $\alpha$-capture to form $^{16}$O, and so on. These additional $\alpha$-capture reactions, which are much more rapid than the 3$\alpha$ reaction \citep{Shen2014A}, will also change the energetic results, and hence the possibility of a self-sustaining flame and subsequent detonation is increased. Furthermore, the analytical curves for He, assuming only $3\alpha$ reactions, show that He below a density $\sim 8.3 \times 10^4$~g~cm$^{-3}$ would not ignite, no matter how high the temperature of the material is. However, this is clearly not the case when one considers the influence of $\alpha$-capture, because above a certain temperature, $\sim 6.02 \times 10^9$~K (see figure \ref{helium46}), He fuses to form $^{12}$C and subsequently higher intermediate mass elements via $\alpha$-capture reactions.\\

We find that as we include more isotopes in our reaction networks, there are more pathways that allow Helium to ignite on much shorter time-scales. There is a cutoff at density $\sim 5 \times 10^4$ gm/cm$^3$ for pure Helium burning which is the characteristic of smaller reaction networks and restricted path ways. However, it is remarkable that even a small amount of carbon mixed with the pure Helium makes the fuel unstable corresponding to a larger fraction $T-\rho$ plane. 
 
In the models presented in \cite{Kashyap2017}, the maximum temperature is found to be just outside the surface of the primary WD. In the spiral mechanism, the hot material at the surface mixes with the cold degenerate material and drives the thermodynamic conditions towards C/O or He ignition. Hence, the investigation of the maximum temperature and density at the location of the maximum temperature is a good indication of the possibility of He ignition in our models. We have found that three lower mass models ($0.8\, M_{\sun} + 0.5 M_{\sun}$,$1.0\, M_{\sun} + 0.6 M_{\sun}$,$1.0\, M_{\sun} + 0.9 M_{\sun}$) are well inside He-ignition region of our evolved He ignition region (see Fig. 8 of \citealt{Kashyap2017}). These findings suggest that white dwarfs with masses in range  $0.8\, M_{\sun}$ to $1.0\, M_{\sun}$ but including thin He surface layers may undergo He-ignition, and possibly would detonate to produce either a `.Ia' or a SN Ia. However, our results suggest that the He ignition would most probably happen during the rapid accretion phase of dynamical merging \citep{Bildsten2007A} because the thermodynamic conditions will reach, for the first time, inside the He-ignition curve during initial phase of dynamical mass transfer.

\subsubsection{Ignition in C/O mixed with trace Helium}
\begin{figure}
	\begin{center}
		\includegraphics[trim=10 0 10 10,clip,width=0.9\columnwidth]{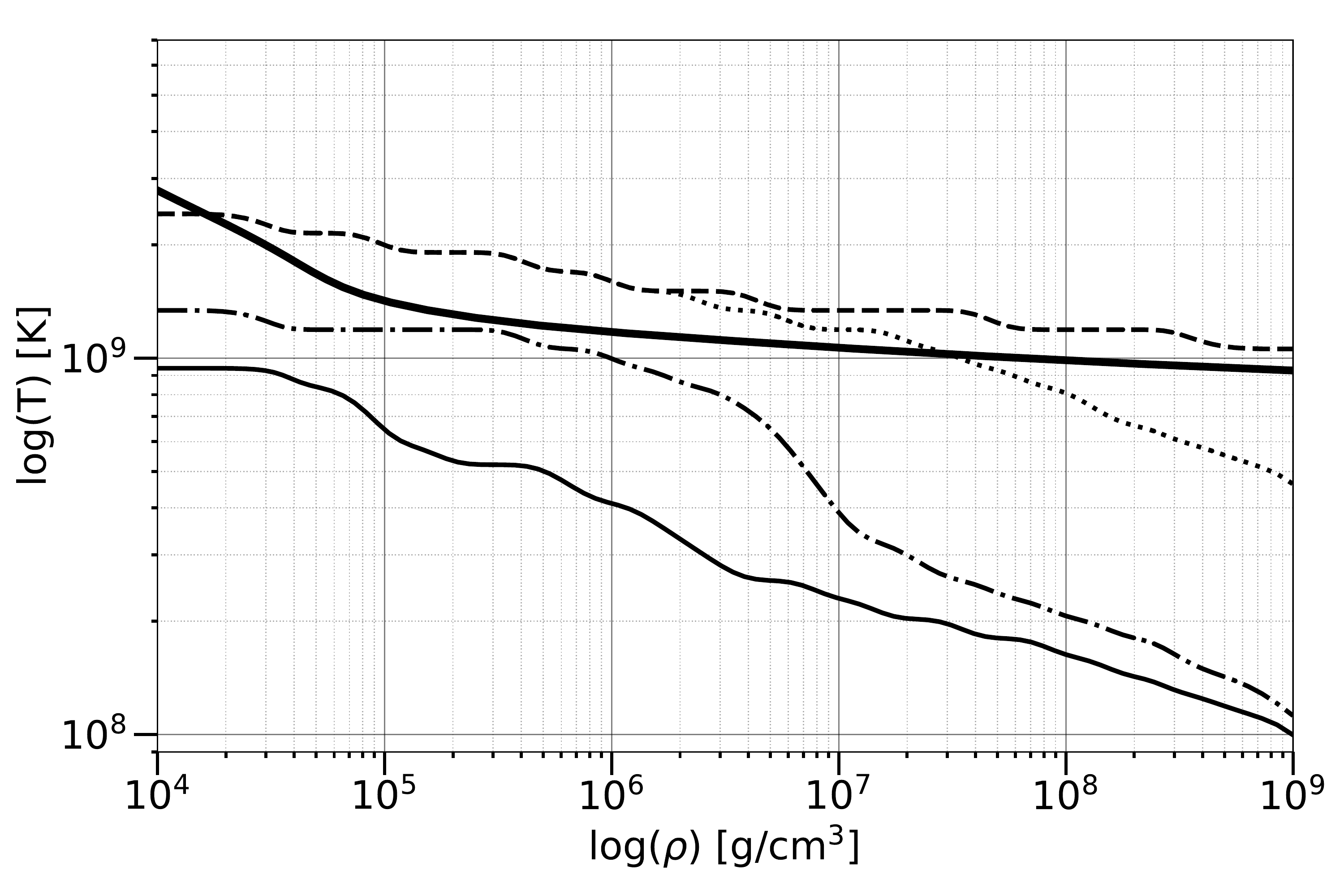}
		\caption{Local ignition criteria in $T-\rho -$plane for primariily-carbon mixture. The thick solid curve is the analytical ignition curve for 50/50 C/O mixture. The dashed (0/50/50), dotted (1/49/50), dot-dashed (5/45/50) and thin solid (20/30/50) black curves are numerical ignition curves using 46 isotope network for different initial initial mixture of He/C/O fuel.}
		\label{carbon46}
	\end{center}
\end{figure}

We plot the unstable ignition curves in figure \ref{carbon46}, along with the analytical $t_{\rm dyn}=t_{\rm nuc}$ curves. In the production of the analytical curves, it has been assumed that all $^{12}$C fuses via $^{12}$C ($^{12}$C,$\alpha$) $^{20}$Ne to form $^{20}$Ne . These assumptions, while are mostly valid for a short time under high temperature conditions, are not fully justified for rapid carbon ignition.\\  

In figure \ref{carbon46}, we present the ignition curves for C/O fuel mixed with varied amount of He. We find that even for 50/50 C/O fuel analytical curve overestimates the conditions for unstable ignition (used as criteria for detonation in the literature). This is mostly because the analytical curve does not take into account the neutrino cooling which allows the curve to go up in $T-\rho$-plane. Mixing the initial fuel with He introduces fast $\alpha$-capture reaction which produces more energy to allow for unstable ignition. This effect is more important in high density region as apparent from figures \ref{helium46} and \ref{carbon46}. This has favorable implications for WDs allowing them to ignite unstably at temperature as low as $2 \times 10^8$ K depending upon the density. The double-detonation scenario has proved to be difficult to obtain in 3D simulations so far. Our results alleviate the chances of this channel as well and widen the chances of SN Ia population via binary sub-Chandrasekhar WDs. Moreover, the mixing of He and C/O in expected in the accretion stream from a degenerate companion to be composed of He \cite{Guillochon2010A}.\\

\section{Conclusion}

We find that large reaction networks may play significant role in He/C/O WD. These effects are especially important for binary sub-M$_{Ch}$ WD mergers. However it is not feasible for them to be included in the hydrodynamical simulations. The results of numerical simulations presented in the results show as well that the enhanced nuclear energy release with very small mixing can broaden the fractions of these binaries that can become SNe Ia. We do not explore the minimum length scales required to start the self-sustained detonation but, we expect them to become smaller in the case of approx46 network compared to \cite{Seitenzahl_etal_2009} (with 13 isotope-network) and \cite{Zenati2020} (with 19 isotope-network for few specific $T-\rho$ points), an issue which has been addressed here. We point out that early indications of results presented here can be guessed from previous studies such as \cite{Shen2014A} where they also identify the key reactions that are responsible for such increase in nuclear heating output. However, they focus only on He burning and handful of temperature-density conditions. In summary, the main conclusions of this paper are as follows:

\begin{enumerate}
\item The local criteria of unstable ignition has been derived for any given fuel in stellar conditions \footnote{The modified version of the nuclear reaction network code,  TORCH and analysis scipt for TORCH output data can be made available to anyone on request to the author.}.  These conditions differs significantly from the analytical calculations, especially for mixed fuels (50/50 C/O with trace He and, pure He with trace C/O repectively) in high temperature regime. The methodology presented here can be used to find the conditions of unstable nuclear burning in accreting stellar remnants of all kinds. Some examples could be surface of neutron stars, white dwarfs and to assess the ignition in silicon, calcium and other heavy metal rich regions in the cores of massive stars. 

\item We have presented the ignition curves using 46 isotope network. The larger network ignition curves are expected to agree with analytical ignition curves in lower temperature regime as many of the reactions used in large reaction network will not be active at lower temperatures. Indeed we find such an agreement for He ignition curves in the regime of $1.5 \times 10^8 < T < 6 \times 10^8$ K and $9 \times 10^4 < \rho < 10^7$ gm/cm$^3$. We obtain a different ignition curve for 50/50 C/O fuel owing most likely to neutrino cooling which has been absent in the calculation of the analytical curves. These curves change the possibility of unstable nuclear burning anticipated previously in literature such as \cite{Dan2014A,Dan2012A}. We have checked that the ignition curves remains almost unchanged when employing more number of isotopes such as 204 isotopes.

\item We find that a fuel mixed with even a small amount of impurity (such as He/C/O = 95/2.5/2.5 or 1/49/50) is more likely to have unstable burning than a pure initial fuel (i.e. He/C/O = 100/00/00 or 00/50/50). A mixture of fuel indeed represents a realistic case involving He layer on the surface of white dwarfs as well as in the accretion stream during dynamical mass-transfer of He from a degenerate companion.

\item Similarly C/O fuel is more likely to detonate when mixed with He (compare thin solid line in Figure \ref{carbon46} and all mixed fuel in Figure \ref{helium46} ). This allows the possibilities for unstable nuclear burning in sub-Chandrasekhar models at lower than the required temperature at lower density for 50/50 C/O fuel often used in literature. 

\end{enumerate}

We have shown that calculations from an extended nuclear reaction network (using 204 species as opposed to 19 species generally used in literature) provides characteristically different conditions for pure He and C ignition, as compared to the analytical results used previously as a criterion of detonation. The effects of turbulent background will allow us to find the implications of results in this paper \citep{Zenati2020} . We emphasize that it is not feasible to include a large reaction network in the full 3D simulations. Hence, a subgrid modelling along with a condition for detonation initiation will be required to capture realistic evolution and correct progenitor scenario. The additional reaction pathways causing the differences in ignition condition are very important as well. We defer such investigations to our future studies. In light of these findings, if the lower mass binary WD models considered in \cite{Kashyap2017} are prone to detonation, they might span a range of luminosities and decline rates covering the full range of normal SNe Ia providing a possible closure to the Type Ia progenitor problem. 

\acknowledgments
I thank all the people who have influenced this work with their suggestions and insights. I would especially like to thank Robert Fisher, James Guillochon, Frank Timmes and Ken Shen for insightful conversations and encouragements.

\software{FLASH, TORCH, yt}

\appendix
\section{List of isotopes}
Our preliminary results demonstrates that a 46-isotope network enables unstable nuclear burning over a wider range of physical conditions as compared to the smaller Approx19 network. The list of isotopes used in 46-isotope network (according to TORCH general network code) is shown in Table~\ref{isotopetable}. In future studies, we plan to identify the reaction pathways that enables the abovementioned changes in larger network. Neutrino losses have been taken into account in our study \citep{Itoh_1996}. 

\begin{table}[htbp]
	\begin{center}
		\centering
		\caption{List of 46 isotopes used in calculation of nuclear reaction network.}
		\vspace{0.5cm}
		\resizebox{0.5\textwidth}{!}{
			\begin{tabular}{|c|c|c|c|}
				\hline 
				Protons&$^2$H&$^3$He&$^4$He\\$^7$Li&$^7$Be&$^8$B&$^{12}$C\\$^{13}$C&$^{13}$N&$^{14}$N&$^{15}$N\\$^{14}$O&$^{15}$O&$^{16}$O&$^{17}$O\\$^{18}$O&$^{17}$F&$^{18}$F&$^{19}$F\\$^{18}$Ne&$^{19}$Ne&$^{20}$Ne&$^{23}$Na\\$^{23}$Mg&$^{24}$Mg&$^{27}$Al&$^{27}$Si\\$^{28}$Si&$^{30}$P&$^{31}$P&$^{31}$S\\$^{32}$S&$^{35}$Cl&$^{36}$Ar&$^{39}$K\\$^{40}$Ca&$^{43}$Sc&$^{44}$Ti&$^{47}$V\\$^{48}$Cr&$^{51}$Mn&$^{52}$Fe&$^{55}$Co\\$^{56}$Ni&Neutrons & & \\
				\hline
		\end{tabular}}
		\label{isotopetable}
	\end{center}
\end{table}

\bibliography{ref_SNIa}{}

\end{document}